\documentclass[12pt]{article}
\pdfoutput=1
\usepackage{graphicx}

\usepackage{amssymb}

\def\fnote#1#2{\begingroup\def\thefootnote{#1}\footnote{#2}\addtocounter{footnote}{-1}\endgroup}

\def\inbar{\vrule height1.5ex width.4pt depth0pt}
\def\IB{\relax{\rm I\kern-.18em B}}
\def\IC{\relax\,\hbox{$\inbar\kern-.3em{\rm C}$}}
\def\ID{\relax{\rm I\kern-.18em D}}
\def\IE{\relax{\rm I\kern-.18em E}}
\def\IF{\relax{\rm I\kern-.18em F}}
\def\IG{\relax\,\hbox{$\inbar\kern-.3em{\rm G}$}}
\def\IH{\relax{\rm I\kern-.18em H}}
\def\II{\relax{\rm I\kern-.18em I}}
\def\IK{\relax{\rm I\kern-.18em K}}
\def\IL{\relax{\rm I\kern-.18em L}}
\def\IM{\relax{\rm I\kern-.18em M}}
\def\IN{\relax{\rm I\kern-.18em N}}
\def\IO{\relax\,\hbox{$\inbar\kern-.3em{\rm O}$}}
\def\IP{\relax{\rm I\kern-.18em P}}
\def\IQ{\relax\,\hbox{$\inbar\kern-.3em{\rm Q}$}}
\def\IR{\relax{\rm I\kern-.18em R}}
\def\IT{\relax{\rm I\kern-.18em T}}
\def\ZZ{\relax{\sf Z\kern-.4em Z}}

\def\a{\alpha}       \def\g{\gamma}  
\def\e{\epsilon} \def\G{\Gamma}   \def\k{\kappa}  
\def\L{\Lambda}     \def\si{\sigma}

\def\cA{{\cal A}}  
  \def\cF{{\cal F}}
 \def\cH{{\cal H}} 
 \def\cK{{\cal K}} 
  \def\cO{{\cal O}}
\def\cP{{\cal P}} \def\cQ{{\cal Q}} \def\cR{{\cal R}}
\def\cS{{\cal S}}

  \def\tf{{\tilde f}}   \def\tg{{\tilde g}}

\def\ttg{{\tilde \tg}}

   \def\bE{{\bar E}}  \def\bF{{\bar F}}

\def\Ewhat{{\widehat E}}

\def\adot{{\dot{a}}}

\def\Hdot{{\dot{H}}}

\def\cRdot{{\dot{\cR}}}   \def\cSdot{{\dot{\cS}}}

\def\phidot{{\dot{\phi}}}   \def\sidot{{\dot{\si}}}

 \def\veck{{\vec{k}}}    \def\vecx{{\vec{x}}}

            \def\rmIm{{\rm Im}}

           \def\rmMpc{{\rm Mpc}}    
\def\rmNL{{\rm NL}}       
          
\def\rmPl{{\rm Pl}}      \def\rmPSL{{\rm PSL}}
    
       \def\rmSL{{\rm SL}}      \def\rmSO{{\rm SO}}

             \def\rmcorr{{\rm corr}}
         
       \def\rmdet{{\rm det}}     
\def\rmdim{{\rm dim}}       
\def\rmdiv{{\rm div}}              \def\rmfl{{\rm fl}}

\def\rmkin{{\rm kin}}         
      \def\rmnmod{{\rm nmod}}   

        \def\rmmod{{\rm mod}}

\def\rmord{{\rm ord}}

\def\rmth{{\rm th}}

\def\rmRe{{\rm Re}}

\def\lfrak{{\mathfrak l}} 
 \def\sfrak{{\mathfrak s}}

 \def\mathC{{\mathbb C}} 
  
  \def\mathR{{\mathbb R}}
\def\mathZ{{\mathbb Z}}

\def\fnote#1#2{\begingroup\def\thefootnote{#1}\footnote{#2}\addtocounter{footnote}{-1}\endgroup}

\def\beq{\begin{equation}}
\def\eeq{\end{equation}}
\def\bea{\begin{eqnarray}}
\def\eea{\end{eqnarray}}
\def\llea#1{\label{#1}\eea}
\def\lleq#1{\label{#1}\eeq}
\let\nn=\nonumber
\def\tabroom{\hbox to0pt{\phantom{\Huge A}\hss}}
\def\notin{\ \hbox{{$\in$}\kern-.51em\hbox{/}}}

\def\lra{\longrightarrow}

\def\del{\partial} \def\vphi{\varphi}

  \def\E1Fq{E_1/\IF_q}

\def\notdiv{{\relax{~|\kern-.34em /~}}}

      \def\oj{{\overline{j}}}
   
\def\ow{{\overline{w}}}

\def\oE{{\overline{E}}}

 \def\otau{{\overline{\tau}}}

\def\boxit#1{
\vbox{\hrule height1pt\hbox{\vrule width1pt\kern0.3cm
\vbox{\kern0.3cm\hbox{$\displaystyle#1$}\kern0.3cm}\kern0.3cm\vrule
width1pt}\hrule height1pt}}

\begin{document}

\phantom{\hfill \today}

\baselineskip=19pt
\parindent=0pt

\vskip .9truein

\centerline{\Large {\bf Modular Inflation Observables and }}

\vskip .2truein

\centerline{\Large {\bf $j$-Inflation Phenomenology}}

 \vskip .4truein

 \centerline{{\sc~ Rolf Schimmrigk$^{\diamond}$}\fnote{}{$\diamond$netahu@yahoo.com,
rschimmr@iusb.edu}}

\vskip .2truein

\centerline{Indiana University South Bend} \centerline{1700
Mishawaka Ave., South Bend, IN 46634}

\vskip 1truein

\centerline{\bf Abstract}
\begin{quote}
  Modular inflation is the restriction to two fields of automorphic inflation,
  a general group based framework for multifield scalar field theories with curved
  target spaces, which can be  parametrized by the comoving curvature
  perturbation $\cR$ and the isocurvature  perturbation tensor $S^{IJ}$.
  This paper describes the dynamics and observables of these perturbations and considers
  in some detail the special case  of modular inflation as an
  extensive class of two-field inflation theories with a conformally flat target space.
  It is shown that the nonmodular nature of derivatives of modular forms leads to
  CMB observables in modular invariant inflation theories that are in general constructed
  from almost holomorphic modular forms. The phenomenology of the model
  of $j$-inflation is compared to the  recent observational  constraints from
  the {\sc Planck} satellite and  the BICEP2/Keck Array data.
\end{quote}

\renewcommand\thepage{}
\newpage
\parindent=0pt

 \pagenumbering{arabic}

\baselineskip=19.5pt
\parskip=.015truein

\tableofcontents

\vfill \eject

\baselineskip=21.7pt
\parskip=.12truein
\parindent=15pt

\section{Introduction}\label{ra_sec1}

Inflation is a framework defined on the space $\cF(M,X)$ of scalar
field theories defined by field multiplets $\phi^I$,
$I=1,...,n$ on the spacetime manifold $M$ taking values in the
target space $X$. The field space $\cF(M,X)$ is in general assumed to
be an unstructured set on which dynamical variables and
observables are defined, providing the underlying theory spaces of
inflation. The dynamics of the background is determined by the
metric $G_{IJ}(\phi^K)$ on the target space and the potential
$V(\phi^I)$. While most of the literature assumes that the target
space is flat, following \cite{s85, sbb89},  $X$ can in general
have a nontrivial geometry, encoded in the Riemannian metric
$G_{IJ}$ that defines the kinetic term
$\frac{1}{2}G_{IJ}g^{\mu\nu} \del_\mu \phi^I\del_\nu\phi^J$.
 Early references that admit a nontrivial geometry
 include \cite{ss95, ns96, gt01, gs02}.
 Depending on the origin of the model, the action can
either be taken to be given by $E$ and $V$ in polynomial form, or
more generally can be a non-rational function, such as in the case of
brane-induced inflation.

There are different reasons why it is of interest to consider a symmetry based
approach to inflation. One motivation arises from the shift-symmetry, an
ad-hoc operation that is often invoked because of concerns about
 higher order corrections to observable parameters
of inflationary models. This is reminiscent of duality considerations, which
have led in gauge and string theory to the embedding of analogous dualities into
larger discrete and continuous groups. A group-theoretic approach
is furthermore useful because it endows the inflaton target space with structure that
allows the exploration of the inflationary theory space in a more systematic
way by using the resulting geometry as an organizing principle.
Symmetries appear as well in inflation theories based on moduli,
sometimes also called modular inflation, in which inflation is driven by some
of the moduli that appear in many fundamental theories \cite{q16}. Such
moduli based models
can provide special cases of the framework developed here.

The idea of automorphic inflation is to consider the space of
the inflationary multifield space defined by theories that are
invariant under infinite discrete symmetry groups that contain the
inflaton shift symmetry. This leads to the notion of automorphic
inflation as a structured framework of multifield inflation
\cite{rs14, rs15}. An immediate consequence of the existence of such
symmetries is that the space of automorphic field theories acquires a
foliation, with leaves that are specified by numerical
characteristics, defined in terms of the group theoretic and automorphic structure,
specifically the underlying continuous group $G$, the discrete group $\G$
in $G$ that extends the shift symmetry, and the types of the automorphic
forms that define the building blocks of the inflaton potential.

 The simplest case of automorphic inflation is obtained when the
 discrete group is the modular group, leading to the special case of
modular inflation, in which case the continuous group $G$ is fixed to be
the M\"obius group $\rmSL(2,\mathR)$ and the discrete group is a
subgroup $\G\subset \rmSL(2,\mathZ)$.\fnote{1}{In the case of inflation
             driven by two or more moduli fields, the presence of appropriate
             symmetries places such models in the framework of modular,
             and more generally, automorphic inflation.}
 The theory of modular forms has
been developed over the past century in greatest
 detail for groups $\G$ that are of congruence type with
 some level $N$, but more general groups of Fuchsian type are also possible.
 The potential $V(\phi^I)$ is constructed in terms of
modular forms $f_i$ of weight $w_i$ relative to $\G(N)$.
 The resulting theory space of modular inflation is thus determined
  by the numerical characteristics $(N,w_i)$.
The kinetic term of all modular inflation models is determined by the
Poincar\'e metric of the upper halfplane, leading to a conformally flat target
space for the inflaton. The invariance of this metric under the continuous group
of M\"obius transformations leads via the introduction of the inflaton potential
 to a breaking of the  group $\rmPSL(2,\mathR)$ to the discrete group
 of modular transformations $\rmPSL(2,\mathZ)$, or subgroups thereof. The energy
 scale of this breaking is determined by CMB constraints, leading to a weakly broken
 symmetry in the case considered here.

Automorphic inflation was briefly outlined in \cite{rs14}, and the general framework
was described in more detail in \cite{rs15}, with emphasis on the automorphic
 side of the theory. The purpose of the present paper is to focus in more detail
 on the class of modular inflation and the particular model of $j$-inflation.
 The phenomenological analysis of general modular inflation
 involves derivatives of the potential, hence derivatives of modular forms. However,
 derivatives of modular forms are not modular forms, raising the issue of the modular
 structure of observables in this framework. It is shown that the basic CMB
 observables are in general determined not by modular forms per se but by
 almost holomorphic modular forms.  The phenomenological analysis of the specific
 model of $j$-inflation, introduced in \cite{rs14}, is compared to observational
 constraints from the microwave anisotropy.  Sections 2 and 3 briefly introduce the general
multifield inflationary background dynamics and perturbation theory, with emphasis on a closed
system of equations for the isocurvature perturbation described by an antisymmetric rank
two tensor $S^{IJ}$. This generalizes an earlier result in the special case of 
flat two-field inflation to an arbitrary
 number of fields with a general curved target space. Sections 4 and 5 consider modular inflation,
 with focus on the general structure of physical observables in this framework, showing in
 particular that they are modular in a generalized sense. Section 6 briefly describes 
 the integration of the inflationary
evolution in terms of the transfer function, and Sections 7 and 8 analyze the
 specific model of $j$-inflation. The final Section presents
some conclusions.

\vskip .4truein

\section{Multifield dynamics}

Automorphic inflation as a group theoretic framework for
multifield inflation involves field spaces that are obtained as coset spaces of
continuous groups, hence are curved.  The specialization to
modular inflation leads to an extensive class of two-field models with
conformally flat target spaces. In this section some
essential features of these theories are described. There are different
ways to encode the dynamics of the adiabatic and isocurvature perturbations.
In the following the comoving curvature perturbation $\cR$ will be adopted as
the adiabatic mode, while the isocurvature perturbations are encoded in an antisymmetric
  tensor denoted by $S^{IJ}$.  The focus on the latter is suggested by the dynamics of $\cR$
  and leads to an isocurvature dynamics different from the usual dynamics based on projections
 of the Sasaki-Mukhanov variables.
 While the geometry in automorphic inflation is derived from the
 structure of the underlying Lie group $G$ and certain subgroups,
 it is best to leave the metric of the  inflaton field space arbitrary
  and the number of fields of the inflaton multiplet $\phi^I$
   unconstrained, so as to indicate the general features of the framework.
Modular inflation is then recovered by setting $G=\rmSL(2,\mathR)$ and
considering the field space $G/K$ with compact subgroup $K=\rmSO(2,\mathR)$,
which is isomorphic to the complex upper halfplane.

\subsection{Scalar field dynamics}

 The scalar field action is considered to be of the
 type
   \beq
   \cA[\phi^I, G_{IJ}, g_{\mu\nu}]
        ~=~  \int d^4x \sqrt{-g} \left(\frac{M_\rmPl^2}{2} R
              -  \frac{1}{2} G_{IJ} g^{\mu \nu} \del_\mu \phi^I \del_\nu \phi^J
              - V(\phi^I) \right),
  \lleq{scalar-action}
  where the spacetime metric has the signature $(-,+,+,+)$ and
   $M_\rmPl = 1/\sqrt{8\pi G_N}$ is the reduced Planck mass associated to
   Newton's constant $G_N$. The target space spanned by the inflaton
   fields $\phi^I$, $I=1,...,N$, in general has a non-trivial geometry
   determined by the metric $G_{IJ}$ that is  assumed to be Riemannian.

The dynamics of the system $(g_{\mu\nu}, G_{IJ}, \phi^I)$ involves
the geometry of the target space as well as that of spacetime via
the Einstein equations and the Klein-Gordon
 equation. Assuming that the covariant derivative on the inflaton space is of
 Levi-Civita type, the Euler-Lagrange form of the latter takes the form
 \beq
 \square_g \phi^I + \G^I_{JK} g^{\mu \nu}
   \del_\mu \phi^J \del_\nu \phi^K - G^{IJ} V_{,J} ~=~0,
 \eeq
 where $\square_g = \frac{1}{\sqrt{-g}} \del_\mu \sqrt{-g} g^{\mu \nu}\del_\nu$ is the
 d'Alembert operator
 and $\G^I_{JK}$ are the target space Christoffel symbols. This Klein-Gordon
 equation can be written
 in terms of a covariant derivative $D_\mu$, which can be viewed as a combination of the
 spacetime Koszul connection
 \beq
 \nabla_{\del_\mu} \del^\k \phi^I ~=~ \del_\mu \del^\k \phi^I + \G_{\mu \nu}^\k \del^\nu \phi^I
 \eeq
 and a contribution of the curved target space
 \beq
 D_\mu \del^\k \phi^I ~=~ \nabla_{\del_\mu} \del^k \phi^I + (\del_\mu \phi^J) \G_{JK}^I \del^\k \phi^K
 \eeq
 as
 \beq
  D_\mu (\del^\mu \phi^I) ~-~ G^{IJ}V_{,J} ~=~ 0.
 \eeq

\subsection{Curved target background dynamics}

The dynamics of the background is assumed to belong to the class
of theories characterized by the action (\ref{scalar-action}). The
 Klein-Gordon equation is given by
 \beq
  D_t \phidot^I ~+~ 3H\phidot^I + G^{IJ}V_{,J} ~=~ 0,
\eeq
 where the dot indicates the derivative with respect to $t$ and
 $D_t$ is the covariant derivative on the target space, defined for vector fields $W^I$
 in terms of the connection coefficients $\G_{IJ}^K$  as
 \beq
 D_t W^I ~=~ \del_t W^I ~+~ \G^I_{JK}\phidot^J W^K.
 \eeq
 Here the $\G_{IJ}^K$ are universally assumed to be the Christoffel symbols determined by the
  Levi-Civita connection. With $\del_t = (\del_t \phi^I) \del_I$ this translates into
 $D_t W^I = \phidot^J \nabla_{\del_J} W^I$, where $\nabla_{\del_I}$ is the Koszul derivative
 of the target space
 \beq
   \nabla_{\del_I} W^J ~=~ \del_IW^J + \G^J_{IK} W^K.
  \eeq

The background equations constraining the  Hubble parameter $H=\adot/a$ are given in the
Newtonian gauge with metric
\beq
  ds^2 ~=~ -dt^2 + a^2(t) \g_{ij} dx^idx^j,
 \eeq
 where in the present paper the spatial metric $\g_{ij}$ is chosen to be flat for simplicity,
by the two Friedman-Lemaitre equations
 \beq
    H^2 ~=~ \frac{\rho}{3M_\rmPl^2}
\eeq
 and
 \beq
    2 \frac{\ddot{a}}{a} ~+~ H^2 ~=~ -\frac{ p}{M_\rmPl^2},
 \eeq
where the density $\rho$ and the pressure $p$ of the isotropic
fluid tensor
 \beq
  T_{\mu\nu} ~=~ (\rho+p)u_\mu u_\nu + p g_{\mu \nu}
 \lleq{fluid-em-tensor}
 are given by
 \bea
  \rho &=& \frac{1}{2}G_{IJ} \phidot^I \phidot^J ~+~ V(\phi^I) \nn \\
  p &=&  \frac{1}{2}G_{IJ} \phidot^I \phidot^J ~-~ V(\phi^I).
 \llea{rho+p}
It is useful to note that the variation of the Hubble parameter is given by
 \beq
  \Hdot
         ~=~ - \frac{1}{2M_\rmPl^2} ~(G_{IJ} \phidot^I \phidot^J).
\eeq
 In all these background equations the functions on field space are  functions
 of the background fields $\phi^I(t)$.

\subsection{Slow-roll dynamics}

In the slow-roll approximation the kinetic energy is assumed to be small compared
to the potential energy,  $\rho_\rmkin \ll V$, and the acceleration of the inflaton is
assumed to be small as well. More precisely, it is conventional to introduce the
parameters
 \bea
  \e &:=& - \frac{\Hdot}{H^2} ~=~ \frac{3\sidot^2}{\sidot^2+2V} \nn \\
  \eta &:=&   \frac{1}{H\e} \frac{d\e}{dt},
 \eea
 where the background dynamics
   has been used and $\sidot \equiv v =  \sqrt{G_{IJ}\phidot^I \phidot^J}$ denotes the background
 inflaton speed, in the notation of \cite{g00etal, pt11}, respectively.
 The parameter $\eta$ can be written as \cite{est12}
 \beq
 \eta ~=~ 2\e ~+~ \frac{2}{H} \frac{\phidot^I
 D_t\phidot_I}{G_{KL}\phidot^K\phidot^L}
  ~=~ 2\e ~+~ \frac{2}{H} \frac{\si^ID_t\phidot_I}{\sidot},
 \eeq
 which shows that the slow-roll approximation constraint $\eta \ll 1$ translates into a
 small projection of the acceleration vector onto the inflaton velocity.

 The slow-roll form of the background
 equation can then be used to eliminate the background field
 velocity
 \beq
   \phidot^I ~=~ - \frac{G^{IJ}V_{,J}}{3H}
  \eeq
 and the slow-roll form of the first Friedman-Lemaitre equation reduces to
 \beq
   H^2 ~=~ \frac{V}{3M_\rmPl^2}.
 \eeq

\vskip .2truein

\section{Perturbed multifield inflation}

While $j$-inflation is an example of two-field inflation, it is
conceptually more transparent to leave the number of fields in the
following brief discussion of inflationary perturbation theory
unrestricted. For scalar field theory with flat target spaces this
has been considered in many references,
  including \cite{s85, sbb89, g00etal, w02etal},
  and reviews can be found in \cite{btw06, w07}.
A comprehensive review for curved target space inflation has
not yet been written, but the references
     \cite{w08, lr08, l08etal, pt11, s12, m12etal, bhp12, kms12,
             ssk13, e14etal, a14etal, g16}
 contain brief descriptions of some
aspects of multifield inflation with a non-trivial field space
geometry, and \cite{s12, est12, m12etal, bhp12, kms12, dfs15,
d15etal} are concerned with covariant formulations that extend the
construction of \cite{gt11}.

The metric perturbations are conventionally parametrized
 as
  \beq
  ds^2 ~=~ -(1+2\vphi) dt^2 ~+~ 2aB_{,i} dx^i dt
    + a^2\left((1-2\psi) \delta_{ij} + 2E_{,ij}\right)dx^i dx^j,
 \lleq{perts-convention}
 where $\vphi = \psi$ in the absence of anisotropic stresses, and
 different gauges are defined via the vanishing of some of these
 perturbations. The notation adopted in (\ref{perts-convention})
  is close to that of the reviews \cite{l10, ahkk14}.

 The inflationary dynamics has been constrained over the past
 two decades by the CMB satellite probes COBE, WMAP and {\sc Planck},
 providing experimental results for some observational variables
 associated to the gravitational and inflaton perturbations
 at the percent level, and non-trivially bounding others.

\subsection{Dynamics of perturbations}

A commonly adopted perturbation is the comoving curvature perturbation,
 defined for general fluids in the Newton gauge as \cite{l80, b80, w08}
 \beq
  \cR ~=~ H\delta u ~-~ \psi,
 \eeq
 where $\psi$ is the spatial metric perturbation and
 $\delta u$ is obtained from the divergence part of the
 energy momentum tensor perturbation
 $\delta T_{0i} = -(\rho + p)\delta u_i$ with
  $\delta u_i = \del_i \delta u + u_i$ as
 \beq
 \delta T_{0i}^\rmdiv = -(\rho+p) \del_i \delta u.
 \eeq
  The above definition is often written in a different form by introducing
  $\delta q = (\rho+p)\delta u$ and writing $\cR$ in terms
 of $\delta q$. In multifield scalar field theory with curved target space
 geometry the perturbation can be expressed in terms
 of the Sasaki-Mukhanov variables \cite{s86, m88}
  \beq
    Q^I ~=~ \cQ^I ~+~ \frac{\phidot^I}{H} \psi,
  \lleq{covariant-MS}
  as
   \beq
    \cR ~=~ -\frac{H}{\sidot} \si_I Q^I,
   \eeq
  where $\si^I$ is the normalized inflaton velocity
   $\si^I = \phidot^I/\sidot$ and $\sidot$ is the inflaton speed
   defined above.  Adopting the notation of \cite{kms12},
  the variables $\cQ^I$  denote the covariant form of
   the field perturbations $\delta \phi^I(t,\vecx) := \phi^I(t,\vecx) - \phi^I(t)$
  defined in terms of the geodesic path between the perturbed field
   $\phi^I(t,\vecx)$ and the background field $\phi^I(t)$ given by
   \cite{gt11}
     \beq
   \delta \phi^I ~=~ \cQ^I - \frac{1}{2}\G^I_{JK}\cQ^J\cQ^K + \cdots.
  \lleq{etaIJ}

 For the time evolution of $\cR$ we find here in terms of rank two tensors the equation
   \beq
   \cRdot ~=~  \frac{H}{\Hdot} \frac{1}{a^2} \Delta \psi
          ~+~ \frac{1}{\sidot} S^{IJ}W_{IJ},
 \eeq
 where
 \beq
   S^{IJ} ~:=~ \frac{H}{\sidot} \left(\si^I Q^J - \si^J Q^I\right),
 \eeq
  and the gradient tensor is defined as
  \beq
  W_{IJ} ~:=~ \si_I V_{,J} - \si_J V_{,I}
  \eeq
This suggests identifying the dimensionless variables $S^{IJ}$ as the
fundamental perturbations that source the large scale
time evolution of $\cR$. They will be referred to as the  rank 2 tensor of
the isocurvature perturbations.

In terms of the isocurvature variables $S^{IJ}$ adopted here as the basic nonadiabatic
perturbations,  the slow-roll approximation of the $\cR$-dynamics on large scales is given by
 \beq
  \cRdot ~=~  -2H \eta_{IK} \si^K \si_J S^{IJ}, \tabroom
 \lleq{sr-1}
 where the slow-roll parameter $\eta_{IJ}$ is defined as
 \beq
  \eta_{IJ} ~:=~ M_\rmPl^2 \frac{V_{;IJ}}{V}.
 \lleq{etaIJ}
  The time evolution of the isocurvature perturbation tensor in this approximation is
  given by
  \beq
 D_tS^{IJ}
  = 2H(\eta_{KL}\si^K\si^L -\e) S^{IJ}
      + H\left( \eta_{KL} G^{K[I} S^{J]L} -
       \frac{\e}{3}M_\rmPl^2 \si^{[I} R^{J]}_{KLM} \si^KS^{LM}
       \right), \tabroom
    \lleq{sr-2}
    where the brackets $U^{[I}V^{J]}$ indicate antisymmetrization without the conventional factor
    of 1/2, and $D_t$ is the covariant derivative acting on the contravariant tensor $S^{IJ}$.
    The differential equations (\ref{sr-1}) and (\ref{sr-2}) form a closed system of evolution equations
    for the comoving curvature perturbation $\cR$ and the isocurvature perturbations $S^{IJ}$ which
    shows how in the general multifield case the latter mix during the evolution and how they couple
    to the curvature tensor.
    Equation (\ref{sr-1}) shows that the adiabatic perturbation remains constant
    on large scales if the vector defined by the slow-roll contraction $\eta_{IJ}\si^J$ of the normalized
    inflaton velocity $\si^I$ is orthogonal to the isocurvature contraction $S^{IJ} \si_J$ of $\si^I$.
    Alternatively one can view the rhs as a quadratic form
    \beq
     D_{IJ} ~:=~ G_{IK} S^{KL}\eta_{LJ}
    \eeq
    defined on the tangent space of the target manifold. The system (\ref{sr-1}), (\ref{sr-2}) generalizes
    the flat target two-field dynamics of ref.  \cite{w02etal} to curved field spaces of arbitrary dimension.

 The power spectrum associated to a dimensionless perturbation, in the following generically denoted by
 $\cO(t,\vecx)$,  is defined in terms of the correlator
 as\
   \beq
 \langle \cO(t,\veck)\cO'(t,\veck')\rangle
  ~=:~ (2\pi)^3 \delta^{(3)}(\veck-\veck') P_{\cO\cO'}(k).
  \lleq{power-spec-def}
  with the associated dimensionless power spectrum defined as
   \beq
    \cP_{\cO\cO} ~:=~ \frac{k^3}{2\pi^2} P_{\cO\cO'}.
  \eeq
  In the present case the variables $\cO,\cO'$ are given by the  perturbations
   $(\cR, S^{IJ})$.

The dynamics of $\cR$ identifies the isocurvature tensor contraction as the essential source
of the large scale behavior of the adiabatic perturbation. We associate to the tensor
$S^{IJ}$ a dimensionless isocurvature scalar such that its power spectrum at horizon crossing
is identical to that of $\cR$ by introducing
 \beq
   \cS ~=~ -\frac{1}{2\a \sidot} S^{IJ}W_{IJ},
 \eeq
 where $\a$ is the absolute value of the normalized acceleration vector $\a^I := D_t \si^I$.
  Per construction the power spectrum $\cP_{\cS\cS}$ of $\cS$ as defined in
  (\ref{power-spec-def})   is identical to that of $P_{\cR\cR}$, given in its dimensionless form by
  \beq
 \cP_{\cR\cR} ~=~ \left(\frac{H}{2\pi}\right)^2 \left(\frac{H}{\sidot}\right)^2,
 \eeq
 while the cross-correlation power vanishes at horizon crossing
   \beq
    \cP_{\cS\cS} ~=~ \cP_{\cR\cR},~~~~ \cP_{\cR\cS} ~=~ 0.
  \eeq

 \subsection{The slow-roll form of the power spectrum}

 The slow-roll approximation of the adiabatic power spectrum $\cP_{\cR\cR}$ can be
 expressed directly either in terms of the potential or in terms of
 the slow-roll parameters $\eta_{IJ}$ introduced above and
  \beq
   \e_I ~=~ M_\rmPl \frac{V_{,I}}{V},
   \lleq{epsilon-I}
   which resolves the parameter $\e$ defined above. Using the resulting
  $
  \sidot^2 = (V/3)G^{IJ}\e_I\e_J
  $
   leads to the power spectrum at horizon crossing as
  \beq
 \cP_{\cR\cR}
   ~~=~~ \frac{1}{12\pi^2 M_\rmPl^4} \frac{V}{G^{IJ}\e_I\e_J}.
  \eeq
 As noted above, $\cP_{\cR\cS}=0$ and $\cP_{\cS\cS}=\cP_{\cR\cR}$ at horizon crossing.

The spectral indices $n_{\cO\cO'}$ are obtained from the power
spectrum $\cP_{\cO\cO'}$ as
 \beq
  n_{\cO\cO'} ~=~ 1 + \frac{d\ln \cP_{\cO\cO}}{d\ln k}.
 \eeq
 The shift by unity is conventional for the adiabatic perturbation, but is not always adopted
 for isocurvature perturbations in the literature.
  The slow-roll form of the power spectrum above then leads to the
  spectral index
 \beq
    n_{\cR\cR} ~=~ 1 - 3G^{IJ} \e_I\e_J + 2
  \frac{\eta_{IJ}\e^I\e^J}{G^{KL}\e_K\e_L}.
\lleq{slow-roll-ssi}
  The constraints on $n_{\cO\cO'}$ obtained by WMAP \cite{h12etal}  and
  {\sc Planck} \cite{planck15-13, planck15-20} therefore restrict the shape of the potentials.

\subsection{The tensor-to-scalar ratio}

Gravitational waves play an important role in constraining the viable part of
the inflationary theory space.  While no primordial signal has been detected,
  satellite probes like WMAP and {\sc Planck} have led to  upper bounds that
  models have to satisfy. Conventionally, these bounds are formulated in
  terms of the tensor-to-scalar ratio $r$, which is constructed from
   the tensor power spectrum  \cite{s79}
 \beq
  \cP_T ~=~ \frac{2}{\pi^2} \left(\frac{H}{M_\rmPl}\right)^2
 \eeq
 and the scalar power spectrum. In multifield inflation different forms for $r$ have
 been considered. A convenient definition defines $r$ as
  the ratio of tensor-to-adiabatic scalar amplitude
 \beq
 r ~:=~ \frac{\cP_T}{\cP_{\cR\cR}},
 \lleq{tensor-ratio}
  which leads to the slow-roll parameter form
   \beq
    r ~=~ 8G^{IJ} \e_I\e_J.
  \lleq{tensor-ratio-mfi}
 The tensor spectral index, defined via the ansatz
  \beq
   \cP_T(k) ~=~ \cP_T(k_p) \left(\frac{k}{k_p}\right)^{n_T},
  \eeq
  can be written in terms of the slow-roll parameters as
 \beq
  n_T ~=~ - G^{IJ}\e_I \e_J,
 \eeq
 leading with eq. (\ref{tensor-ratio-mfi}) to the $(r,n_T)$-relation
  \beq
    r ~=~ -8n_T
  \lleq{r-nt-relation}
  in the slow-roll approximation. This relation is affected by the transfer functions, as discussed
  further below.

 \vskip .3truein

\section{Modular inflation}

In this section the framework of automorphic inflation is
specialized to the case of modular inflation, a particular class
of two-field scalar field theories coupled to gravity with a
nontrivial target space geometry, which is of coset type $G/K$, where $G$
is a Lie group and $K\subset G$ is a maximal compact subgroup. The general
framework was introduced in \cite{rs14} and its structure described in
more detail in \cite{rs15} in the higher rank case.

 Classical modular forms \cite{gs73, ds06}
  were introduced in the second half of the 19$^\rmth$ century as functions
 on the complex upper halfplane because this space is mapped to itself by the modular group
 $\rmSL(2,\mathZ)$. The general concept was introduced by Klein \cite{k1890} in the context
 of various discrete  subgroups $\G \subset \rmSL(2,\mathZ)$.
  Thinking about forms in this way is computationally useful, but conceptually
 not the most illuminating approach, and it is most advantageous to have both the domain theoretic
 and the group theoretic formulations available. Such a framework is described in
 refs. \cite{rs14, rs15}.  The result is that in the modular context the group theoretic set-up is given by the pairs of
 group $(G, \G)$, where  $G=\rmSL(2,\mathR)$ is semisimple. The domain theoretic
 structure is obtained by considering a maximal compact subgroup $K\subset G$, which in
 this case is the rotation group $K=\rmSO(2,\mathR)$, both of which act via the
  M\"obius transformation. More details can be found in \cite{rs15}.

 The discrete groups can be of Fuchsian type
 but the most well-developed theory is that of different types
 of congruence groups $\G_N \subset \rmSL(2,\mathZ)$, where the level $N$ determines the
 congruence constraint. For Hecke groups the matrices
 \beq
  \g ~=~ \left(\matrix{
            a &b \cr c &d\cr
             }\right) \in \G_0(N) \subset \rmSL(2,\mathZ)
  \eeq
   satisfy the constraint $c\equiv 0(\rmmod~N)$. Other possibilities include groups usually
 denoted by $\G_1(N)$ and $\G(N)$. These will collectively be denoted as $\G_N$ in the following.

\subsection{Kinetic term}

The bounded domain $\cH$ is two-dimensional, which implies via
the Bianchi identity that the Riemann curvature tensor takes the
 form
 \beq
 R_{IJKL} ~=~ \cK (G_{IK}G_{JL} - G_{IL}G_{JK}),
 \eeq
 where $\cK=R/2$ is the Gaussian curvature expressed in terms of the Ricci scalar $R$.
The metric on $\cH$ induced by the Cartan-Killing
 form $B$ on the Lie algebra $\sfrak\lfrak(2,\mathR)$ is the Poincar\'e metric
 \beq
  ds^2 ~=~ \frac{d\tau d\otau}{(\rmIm~\tau)^2} ~=~
  \frac{dx^2+dy^2}{y^2}.
 \lleq{poincare-metric}
 The field theoretic form of the Poincar\'e metric
 \beq
  ds^2 ~=~ G_{IJ} d\phi^I d\phi^J ~=~ \frac{\mu^2}{(\phi^2)^2}
  \delta_{IJ} d\phi^I d\phi^J
 \eeq
 leads to the non-vanishing Christoffel symbols
 \beq
  \G^2_{11} ~=~ -\G^2_{22} ~=~ - \G^1_{12}
   ~=~ \frac{1}{\phi^2}~=~ \frac{1}{\mu (\rmIm~\tau)}.
 \eeq
 The curvature tensor $R_{IJKL}$ has only one independent
 component
  \beq
    R_{1212}
      ~=~ -  \frac{1}{\mu^2 (\rmIm~\tau)^4}
  \eeq
  leading to the curvature scalar
  \beq
  R ~=~ 2G^{11} G^{22} R_{1212}
 ~=~ -\frac{2}{\mu^2}.
 \eeq
Thus the space $(\cH, ds^2)$ with the metric (\ref{poincare-metric}) has constant negative
Gaussian curvature $\cK = -1/\mu^2$.

The kinetic term, written in multifield notation as
  $G_{IJ} g^{\mu \nu} \del_\mu \phi^I \del_\nu \phi^J/2$
 with $I,J=1,2$  is invariant under the  group defined by the linear fractional transformations
  \beq
   g\tau ~=~ \frac{a\tau+b}{c\tau+d},
   \lleq{moebius-action}
   where
   \beq
   g = \left(\matrix{a&b\cr c&d\cr}\right), ~~~\rmdet~g>0
  \eeq
  are matrices with real entries.
  These M\"obius transformations map the upper halfplane to itself and  define the
  continuous symmetry group of the free modular theory.
  Since the center of $\rmSL(2,\mathR)$ acts trivially the continuous invariance group is
  given by $\rmPSL(2,\mathR)$.

\subsection{Modular potentials and symmetry breaking}

The action of modular inflation can be written as
 \beq
 \cA_\rmmod ~=~ \int d^4x\sqrt{-g} \left(\frac{M_\rmPl^2}{2} R
   - \frac{1}{2}G_{\tau\otau} g^{\mu \nu}\del_\mu \tau \del_\nu\otau
   - V(F(\tau)) \right),
 \eeq
 where $G_{\tau\otau}$ is the complex form of the Poincar\'e metric.
The potentials of modular two-field theory are defined in terms of modular forms
that descend from the group to the upper halfplane. The general construction
of the inflaton target space $X$ in terms of the Lie group $G$ has been described
in detail in \cite{rs15}.

 Modular forms on the upper halfplane
 $\cH$ are induced by the group function $\Phi: \rmSL(2,\mathR) \lra \mathC$
 via the  1-cocycle
 \beq
  J(g,\tau) ~=~ (c\tau + d), ~~~g=\left(\matrix{a&b\cr c&d\cr}\right)
 \eeq
 as
 \beq
   f(\tau) ~=~ (c\tau + d)^w \Phi(g),
 \eeq
 where $\tau = gi$. They are defined  with respect to discrete subgroups $\G_N$ of the
modular group $G(\mathZ) = \rmSL(2,\mathZ)$ and are characterized
by the level $N$ of the subgroups $\G_N$, their weights $w$, and by
 a character $\e_N$ via their transformation behavior,
 which for  $\g \in \G_N$ and the action given by the discrete M\"obius transformation
 $\g \tau = (a\tau+b)/c\tau + d)$ is defined to be
 \beq
 f(\g \tau) ~=~ \e_N(d) (c\tau+d)^w f(\tau).
 \eeq

The potential $V(\phi^I)$ is defined in terms of modular functions $F(\tau)$ on the upper halfplane, as
well as some function $\Phi(F)$ as
\beq
 V ~=~ \L^4 \Phi(F,\bF),
\eeq
where $\L^4$ is an energy scale and $F$ and $\Phi$ are dimensionless.
A simple class of function $\Phi$ is given by powers of the norm function, leading to
 \beq
  V_p ~:=~ \L^4 |F(\tau)|^{2p}.
 \eeq
 In the present paper the focus will be on the $p=1$ case.
 The modular functions $F$ can in general be viewed as a discrete subgroup of the modular group $\rmSL(2,\mathZ)$,
 but the most detailed theory of forms has been formulated for congruence group $\G(N)$ of various types with level $N$.
 The introduction of the potential thus breaks the continuous M\"obius group of the previous subsection to the discrete
 subgroups, which can be written schematically as
  \beq
   \rmPSL(2,\mathR) ~~\lra ~~ \G(N) ~\subset ~ \rmPSL(2,\mathZ).
  \eeq
  In $j$-inflation, the example considered further below, the M\"obius symmetry is weakly broken since the
  constraints from the CMB determine the energy scale
  $\L$ to be much lower than the Planck scale.

\subsection{Modular Eisenstein series}

As in the automorphic case, Eisenstein series play a key role for
modular forms of arbitrary weight because for the full modular
group they span the subspace complementary to the
 cusp forms. Holomorphic Eisenstein series are obtained by following the general construction
 briefly outlined in the general case in the previous section. The details of how to obtain
 from the group theoretic Eisenstein series on $G=\rmSL(2,\mathR)$  the classical Eisenstein
 functions $E_w(\tau)$ on the upper halfplane, given in terms of the divisor function
 \beq
  \si_w(n):=\sum_{d|n}d^w,
 \eeq
 the Bernoulli numbers $B_w$, and  $q=e^{2\pi i \tau}$ with $\tau \in \cH$, as
 \beq
  E_w(\tau) ~=~ 1 -\frac{2w}{B_w} \sum_n \si_{w-1}(n) q^n,
 \lleq{eisenstein-series}
 can be found in ref. \cite{rs15}.

\parindent=20pt
 There are different ways to obtain the values for $B_w$,
 for example via the generating function $x/(e^x-1) = \sum_{m=0}^\infty B_mx^m/m!$,
 or in terms of the Riemann zeta function via Euler's formula as
 $B_w = - 2w! \zeta(w)/(2\pi i)^w$.  For $w>2$ these functions are modular.
  In the case of $j$-inflation the forms of weight $2,4,6$ are relevant and with
   Euler's results for $\zeta(2), \zeta(4)$ and $\zeta(6)$ \cite{e1735}
   the zeta function relation leads to
  $ B_2=1/6, B_4= -1/30, B_6=1/42$.
These ingredients will be used below to define $j$-inflation.

\vskip .3truein

\section{Observables in modular inflation}

It is shown in this section that for general modular invariant inflation the physical observables
are determined by modular forms that are almost holomorphic, but in general not holomorphic.
The explicit form of the spectral index and the tensor-to-scalar ratio
 is determined in terms of the defining modular function $F$ of the inflationary potential.

For two-field inflation the dynamical system introduced above simplifies
considerably because there is only one independent isocurvature perturbation.
The general adiabatic equation (\ref{sr-1}) takes the form
  \beq
   \cRdot ~=~ - 2H \si^K \eta_{K[1}\si_{2]} S^{12},
  \eeq
  which in the case of modular inflation simplifies further because the metric
  is conformally flat $G_{11}=G_{22}=G$, resulting in
  \beq
   \cRdot ~=~ 2HG \eta_{\si s} S^{12}
  \eeq
  with
  \beq
   \eta_{\si s}
   ~=~  (\eta_{22}-\eta_{11}) \si^1\si^2 +  \eta_{12} \left((\si^1)^2- (\si^2)^2\right).
  \eeq

  The isocurvature equation (\ref{sr-2}) reduces to
    \beq
    D_t S^{12}
     ~=~ H\left(\eta_{\si\si} - \eta_{ss} + 2\e \left(\frac{1}{3}M_\rmPl^2 \cK - 1\right)\right)
      S^{12},
  \lleq{isocurvature-sr-dyn}
  where the abbreviations $\eta_{\si\si}= \eta_{IJ}\si^I\si^J$ and
  \beq
   \eta_{ss} ~=~   \eta_{11}(\si^2)^2 - 2\eta_{12} \si^1 \si^2 + \eta_{22} (\si^1)^2
  \eeq
 are covariant objects obtained from  (\ref{etaIJ}).  This specialization of the general dynamics
 of the system $(\cR,S^{IJ}$)  derived above extends the discussion of
  ref. \cite{w02etal} for two-field  inflation with a flat field space metric   to curved target spaces.

 \subsection{Modular inflation parameters}

The inflationary analysis considered in Section 3 involves the
geometry of the potential.  General expressions for the
observables in modular inflation thus involve derivatives of
modular forms. For a potential of modular functions
 \beq
 V ~=~ \L^4 |F|^{2}
 \eeq
 for a modular function $F$ the slow-roll parameters $\e_I$ defined in eq. (\ref{epsilon-I})
 take the form
 \beq
  \e_I ~=~ i^{I-1} \frac{M_\rmPl}{\mu}
   \left(\frac{F'}{F} ~+~ (-1)^{I-1}\frac{\bF'}{\bF}\right),
  \eeq
 and the acceleration of the scale parameter is directly
 determined by the behavior of
  \beq
   \e_V ~:=~ \frac{1}{2}G^{IJ}\e_I\e_J
      ~=~ 2 \frac{M_\rmPl^2}{\mu^2}
        (\rmIm~\tau)^2\left|\frac{F'}{F}\right|^2.
  \eeq
 These parameters determine the tensor-to-adiabatic scalar ratio
  and part of the spectral index.

  The remaining ingredient of  $n_{\cR\cR}$ is the parameter matrix $\eta_{IJ}$
  defined in (\ref{etaIJ}). Decomposing the covariant derivative into its flat and
  Christoffel contributions
  \beq
   \eta_{IJ} ~=~ \eta_{IJ}^\rmfl ~+~ \eta_{IJ}^\G
  \eeq
  leads to the flat part
  \beq
  \eta_{IJ}^\rmfl
  ~=~  -i^{I+J}\frac{M_\rmPl^2}{\mu^2}
         \left( \frac{F''}{F}
           - \left( (-1)^I+ (-1)^J\right) \left|\frac{F'}{F}\right|^2
         + (-1)^{I+J}  \frac{\bF''}{\bF}\right),
  \eeq
 while the Christoffel term $\eta_{IJ}^\G = - M_\rmPl \G^K_{IJ}\e_K$  is given by
  \beq
   \eta_{\IJ}^\G ~=~  -\frac{M_\rmPl^2}{\mu^2}
       \frac{i^{I+J-1}}{\rmIm~\tau} \left(\frac{F'}{F} + (-1)^{I+J-1} \frac{\bF'}{\bF}\right).
  \eeq

 \parindent=0pt
For modular inflation models in which the curvature contribution is small, like
in $j$-inflation, this limit provides a very good approximation to the full result.

\parindent=15pt

\subsection{Modular building blocks of physical observables}

Modular functions $F$ can be written as quotients of modular forms,
hence the computation of $F'$ reduces to the computation of $f'$
for modular forms of some arbitrary weight $w$. However, the derivative
of a modular form is not a modular form. This raises the issue of what precisely
the modular structure is of the physical observables in modular invariant inflation.
The general
structure of the derivative takes the form
 \beq
  f' \equiv \frac{df}{d\tau}
    ~=~ 2\pi i \left(\tf + \frac{w}{12}fE_2\right),
 \lleq{modular-form-derivative}
 where $\tf$ is a modular form of weight $(w+2)$. The problem to
 specify the first term in $f'$ is nontrivial and was described in some detail in \cite{rs15}.
 Very briefly, in the case of the full modular group it is determined by
 functions $H_\tau(z)$ considered in \cite{akn97}
 \beq
  H_\tau(z) ~:=~ \sum_{n=0}^\infty j_n(\tau)q^n,
 \eeq
 where $q=e^{2\pi i z}$ and the functions $j_n$ are constructed iteratively
 via the normalized weight zero Hecke operators $T_0(m)$ as
 $j_n(z) ~=~ j_1(z){\Big |}_{T_0(n)}$,  where $j_1(z) = j(z) - 744$. With
 \beq
  e_\tau ~:=~ \left\{\begin{tabular}{c l}
     $1/2$   &if $\tau =i$ \\
     $1/3$   &if $\tau=\xi_3$ \\
     1   &otherwise \\
     \end{tabular}
     \right\}
  \eeq
   the modular form $\tf$ can be written as
 \beq
  \tf(z) ~=~ -f \sum_{\tau \in \cF} e_\tau \rmord_\tau(f) H_\tau(z),
  \eeq
 where $\rmord_\tau(f)$ is the vanishing order of $f$, which is
 constrained by the valence formula
  \beq
   \frac{1}{2}\rmord_{\tau=i}(f) ~+~
   \frac{1}{3}\rmord_{\tau=\xi_3}(f) ~+~
   \rmord_\infty (f)
    + \sum_{\tau\in \cF-{i,\xi_3}}  \rmord_\tau(f) ~=~ \frac{w}{12},
  \lleq{valence-formula}
  where $\xi_3=e^{2\pi i/3}$ \cite{kk98, o04}.

The nonmodularity of $f'$ arises from the second term because the Eisenstein series $E_2$,
defined as in (\ref{eisenstein-series}), transforms under the modular group $\rmSL(2,\mathZ)$
as
\beq
  E_2(\g \tau) ~=~ (c\tau + d)^2E_2(\tau) - \frac{6i}{\pi} c (c\tau+d),
 \eeq
 and therefore is not a modular form.  This shows that in the inflationary context the relevant
 space is not just generated by $E_4$ and $E_6$, but must also include the Eisenstein series of
 weight two.  $E_2$ is the paradigmatic example of a quasimodular form,
  a notion for which various definitions have been introduced. From a physical point of view
  it is best to focus on the
  transformation behavior of these functions under the modular group.
  The example of $E_2$ indicates a structure that is reminiscent of the case of Christoffel symbols,
  objects with an inhomogeneous transformation behavior, where
  the tensor behavior is modified by an additional term. Similarly, a quasimodular form transforms
  like an ordinary modular form, but with additional terms. For a quasimodular form $f^q$ of weight $w$
  this can be written as
   \beq
    f^q(\g\tau) ~=~ (cz+d)^wf(\tau) ~+~ (cz+d)^w \sum_{m\geq 1} \left(\frac{c}{(c\tau+d)}\right)^m f_m,
  \eeq
  where the sum is finite, and $f_m$ are holomorphic functions.  
  This definition is more direct, but not
  less general, than Nahm's definition as given in \cite{z08}. 
  The original definition of
  quasimodular forms was based on the
   "constant" term of a nearly holomorphic, or almost holomorphic,
  form as defined by Shimura \cite{gs86, kz95}.

  The nonmodularity of $f'$ a priori induces nonmodular terms in the observables of modular inflation,
  hence this raises the question  what exactly the modular nature of these objects is in modular invariant
  inflation.  For modular functions $F=f/g$ with modular forms $f,g$ of equal weight the derivative turns
 out to be simple because of a cancellation and we obtain
  \beq
   \frac{F'}{F} ~=~ 2\pi i \left(\frac{\tf}{f} - \frac{\tg}{g}\right),
  \eeq
  where $\tf, \tg$ are as in (\ref{modular-form-derivative}).
  Thus the first derivative of modular functions are modular forms of weight two. The second derivative $F''$ is
  no longer modular and by using the above formula for the derivative iteratively we find
  \beq
  F'' ~=~ -4\pi i F' \frac{\tg}{g} - 4\pi^2 F \left(\frac{\tilde{\tf~}}{f} - \frac{\tilde{\tg}}{g}\right)
   + \frac{\pi i}{3} F' E_2,
 \eeq
 where $\tilde{\tf~}$  and $\ttg$ are obtained from the derivatives of the modular forms $\tf$ and $\tg$.
 Important for the $\eta_{IJ}$-induced terms in the spectral observables is the combination $F''/F'$. The
 nonmodular part of this quotient is therefore given by
  \beq
   \left(\frac{F''}{F'}\right)_\rmnmod ~=~ \frac{\pi i}{3} E_2.
  \eeq

The main result of this and the following discussion is that there are further terms in the modular inflation
observables that are induced by  the nontrivial geometry of the target space. These terms combine
with the nonmodular Eisenstein series $E_2$ into a
 new function that is modular, but not holomorphic, defined as
  \beq
   \Ewhat_2(\tau) ~=~ E_2(\tau) ~-~ \frac{3}{\pi (\rmIm~\tau)}.
  \eeq
  The modularity of this function follows from the transformation behavior
   \beq
  \frac{1}{\rmIm~\g \tau} ~=~ \frac{(c\tau+d)^2}{\rmIm~\tau} - 2ic(c\tau+d)
    ~=~ \frac{|c\tau+d|^2}{\rmIm~\tau}.
 \eeq

 The Eisenstein series $\Ewhat_2$ is an example of a nearly holomorphic, or almost holomorphic,
 modular form.  Such  a form of weight $w$ for  the modular group $\rmSL(2,\mathZ)$ is a
 function $f$ on the upper halfplane that is a polynomial in $1/(\rmIm~\tau)$ with coefficients that are holomorphic
 functions.

 \subsection{The almost holomorphic modularity of CMB observables}

  The results above can now be used to address the question raised earlier about the
  modular structure of the CMB observables. The tensor-to-scalar ration $r$ is determined by the
 modular form $F'/F$ as
  \beq
   r ~=~ 32 \frac{M_\rmPl^2}{\mu^2} (\rmIm~\tau)^2 \left|\frac{F'}{F}\right|^2,
  \eeq
  hence is modular invariant in terms of holomorphic modular forms (and their complex conjugates).
 The spectral indices involve the second derivatives $F''$, hence are not modular in the same sense.
 From the expressions obtained above for the parameters $\e_I$ and $\eta_{IJ}$
  we obtain for modular inflation the spectral index $n_{\cR\cR}$ as
 \beq
   n_{\cR\cR}
   ~=~ 1 -  4 \frac{M_\rmPl^2}{\mu^2}  (\rmIm~\tau)^2
    \left[ 2 \left|\frac{F'}{F}\right|^2
     ~-~ \rmRe \left(\frac{F''}{F'}  \cdot \frac{\bF'}{\bF}\right) \right]
     ~ - ~4\frac{M_\rmPl^2}{\mu^2} (\rmIm~\tau) \rmIm\left(\frac{F'}{F}\right),
  \lleq{nRRsr}
  where the final term is induced by the target space metric.
 The nonmodular contribution to the second term in the square brackets combines
 with the last term in this equation so that the modularity of the spectral index
 can be made manifest by writing
 \beq
   n_{\cR\cR}
   ~=~ 1 -  4 \frac{M_\rmPl^2}{\mu^2}  (\rmIm~\tau)^2
    \left[ 2 \left|\frac{F'}{F}\right|^2
     ~-~ \rmRe \left(\frac{F''}{F'}  \cdot \frac{\bF'}{\bF}\right)_\rmmod
     ~ + ~\frac{\pi}{3} \rmIm\left(\Ewhat_2 \frac{\bF'}{\bF}\right)\right].
  \lleq{nRRsr-modular}
  This completes the derivation of the most important observables for the general
  framework of modular inflation based on arbitrary modular invariant functions $F$.
The weight structure of the inflationary variables
 can be resolved into their holomorphic and
anti-holomorphic factors $(w,\ow)$. It follows from the above that
the slow-roll parameters $\e_I$ are determined by forms of weight
(2,0) and (0,2), while the parameter $\e_V$, which is proportional
to $r$, is of weight (0,0).

The two-field dynamics at the beginning of this section can now be made explicit  in terms of
the defining modular invariant function $F$ as
 \bea
  \eta_{\si\si}
     &=&  2 \frac{M_\rmPl^2}{\mu^2}  (\rmIm~\tau)^2
       \left[ \left|\frac{F'}{F}\right|^2  ~+~  \rmRe \left(\frac{F''}{F'}  \frac{\bF'}{\bF}\right)_\rmmod
          - \frac{\pi}{3} \rmIm \left( \Ewhat_2  \frac{\bF'}{\bF}\right) \right]
       \nn \\
   \eta_{ss}
       &=&  2 \frac{M_\rmPl^2}{\mu^2}  (\rmIm~\tau)^2
   \left[ \left|\frac{F'}{F}\right|^2  ~-~  \rmRe \left(\frac{F''}{F'}  \frac{\bF'}{\bF}\right)_\rmmod
          + \frac{\pi}{3} \rmIm \left( \Ewhat_2  \frac{\bF'}{\bF}\right)   \right] \nn \\
  \eta_{\si s}
      &=&  -2\frac{M_\rmPl^2}{\mu^2} ~(\rmIm~\tau)^2
   \left[\rmIm\left(\frac{F''}{F'} \frac{\bF'}{\bF}\right)_\rmmod
     ~+~ \frac{\pi}{3} \rmRe \left(\Ewhat_2 \frac{\bF'}{\bF}\right) \right].
 \llea{etas-in-modular-inflation}
 These parameters complete the specification of the two-field dynamics for general modular
 inflation and determine the transfer functions considered in the next section.

  \vskip .3truein

\section{Dynamics via transfer functions}

It is useful to note that a system of equations of the form
 \bea
  \cRdot &=& AH \cS \nn \\
  \cSdot &=& BH \cS
 \eea
 can be integrated formally in terms of transfer functions as
  \bea
   \cR(t) &=& \cR(t_*)+ T_{\cR\cS} \cS(t_*) \nn \\
   \cS(t) &=& T_{\cS\cS} \cS(t_*),
 \eea
 where
  \bea
   T_{\cS\cS}(t,t_*)
     &=& \exp \left(\int_{t_*}^t  dt' B(t')H(t') \right) \nn \\
   T_{\cR\cS}(t,t_*)
     &=& \int_{t_*}^t dt' A(t') H(t') T_{\cS\cS}(t',t_*).
  \eea

  The details of the coefficient functions $A,B$ depend on whether the target space is flat or curved,
  and on the details of the model. For flat targets they can be expressed in terms of the contractions
  of the slow-roll parameters $\e_I$ and $\eta_{IJ}$, as illustrated in the case of two-field
  inflation in \cite{w02etal}. For curved targets the dynamics involves the curvature of the field
  manifold, as indicated by  eq. (\ref{sr-2}). For two-field inflation this specializes to eq.
  (\ref{isocurvature-sr-dyn}),  with parameters given in eq. (\ref{etas-in-modular-inflation}).

The resulting power spectra of the adiabatic and isocurvature perturbations
 \bea
  \cP_{\cR\cR}(t) &=& (1+T_{\cR\cS}^2) \cP_{\cR\cR}(t_*) \nn \\
  \cP_{\cR\cS}(t) &=& T_{\cR\cS}T_{\cS\cS} P_{\cR\cR}(t_*) \nn \\
  \cP_{\cS\cS}(t) &=& T_{\cS\cS}^2 \cP_{\cR\cR}(t_*)
  \eea
  lead to evolving spectral indices
  \bea
   n_{\cR\cR}(t) &=& n_{\cR\cR}(t_*) - (A_*+B_*T_{\cR\cS}) \frac{2T_{\cR\cS}}{1+T_{\cR\cS}^2} \nn \\
   n_{\cR\cS}(t) &=& n_{\cR\cR}(t_*) - \frac{A_*}{T_{\cR\cS}} - 2B_* \nn \\
   n_{\cS\cS}(t) &=& n_{\cR\cR}(t_*) - 2B_*,
  \eea
  where the $*$ on $A,B$ indicates evaluation at horizon crossing.
These indices can alternatively be expressed in terms of the correlation fraction
  $\g_\rmcorr = \cP_{\cR\cS}/\sqrt{\cP_{\cR\cR}\cP_{\cS\cS}} = T_{\cR\cS}/\sqrt{1+T_{\cR\cS}^2}$
  for which constraints have been determined by the {\sc Planck} collaboration.
 The results of eq. (\ref{etas-in-modular-inflation})  complete  the specification of
  the evolution of these spectral indices for general modular inflation.

As mentioned earlier, the
 gravitational tensor power spectrum can be quantified in a variety of ways when
isocurvature perturbations are present. Using the tensor-to-adiabatic scalar power ratio
(\ref{tensor-ratio})  leads via the evolution of $\cP_{\cR\cR}$ to the evolution  of $r$ as
 \beq
  r~=~ \frac{r_*}{1+T_{\cR\cS}^2},
 \eeq
 i.e. to a suppression of $r$ for post-horizon crossing times, hence changing the relation
 between $r$ and $n_T$. Alternatively, this can again be expressed in terms of the
 correlation fraction $\g_\rmcorr$. For more than two fields the adiabatic dynamics
  (\ref{sr-1}) shows that the isocurvature correlators $\langle S^{IJ}(\veck)S^{KL}(\veck')\rangle$
  lead to a further suppression of the tensor ratio, turning the above relation again into
  an inequality, a fact that was anticipated in
  \cite{w02etal} on the basis of a two-field discussion.

 \vskip .3truein

\section{$j$-Inflation}

There are a number of prominent modular forms that can be used
within the framework of modular inflation. The model considered in \cite{rs14}
 is based on a modular function that is basic to all modular forms, the $j-$function, which up to scaling
is the Klein invariant $J(\tau)$.  The fundamental nature of this function is indicated by its
 origin in the Eisenstein series $E_4$ and $E_6$, which
  provide a basis of the space of all modular forms of the full modular group $G(\mathZ) =
\rmSL(2,\mathZ) \subset \rmSL(2,\mathR)$. Denoting the space of
such modular forms by
 \beq
 M_*(\rmSL(2,\mathZ)) ~=~ \bigoplus_w M_w(\rmSL(2,\mathZ)),
 \eeq
 where $M_w(\rmSL(2,\mathZ))$ denotes the spaces of weight $w$ forms, gives
 \beq
 M_*(\rmSL(2,\mathZ)) ~=~ \langle E_4, ~ E_6\rangle,
 \eeq
where the Eisenstein series $E_w$ are defined in (\ref{eisenstein-series}).

 Modular invariant functions with respect to subgroups $\G_N
 \subset \rmSL(2,\mathZ)$ can be constructed by considering
 quotients of modular forms of equal weight. For the full modular
 group the dimension of the space $M_w(\rmSL(2,\mathZ))$ for even $w$ is
 given by \cite{kk98}
 \beq
  \rmdim~M_w(\rmSL(2,\mathZ)) ~=~
    \left\{ \begin{tabular}{l l}
            $\left[\frac{w}{12}\right]$  &for $w\equiv 2(\rmmod~12)$ \\

            $1+\left[\frac{w}{12}\right]$  &for $w\neq
            2(\rmmod~12)$ \tabroom \\
            \end{tabular}
            \right\}.
  \eeq

This shows that for even $w<12$ these spaces are at most one-dimensional.
 At weight $w=12$ one encounters the first cusp form, given by the
 Ramanujan form, which can be written in terms of the
  Dedekind eta function
 \beq
 \eta(\tau) = q^{1/24} \prod_{n=1}^\infty (1-q^n),
 \eeq
 which is closely related to the partition function, hence the
 harmonic oscillator, as
 \beq
 \Delta(\tau) ~=~ \eta(\tau)^{24}~=~ q\prod_{n=1}^\infty
 (1-q^n)^{24},
 \eeq
 or alternatively in terms of the Eisenstein series (\ref{eisenstein-series}) as
 \beq
 \Delta(\tau) ~=~ \frac{E_4^3(\tau)-E_6^2(\tau)}{1728}.
 \eeq
 This form does not vanish on $\cH$, hence one can obtain modular
 functions without poles on $\cH$ by using $\Delta$ as the
 denominator.

 A prominent modular invariant function obtained in this way is
  the  absolute modular invariant defined by the $j$-function
 \beq
 j(\tau)~:=~ \frac{E_4^3(\tau)}{\Delta(\tau)}.
\eeq
 Up to a factor $j(\tau)$ is the Klein invariant $J(\tau) = j(\tau)/1728$ \cite{k1879}.
 The different normalizations are motivated by the fact that the Fourier expansion of
 $j(\tau)$ has integral coefficients
 \beq
 j(q) ~=~ \frac{1}{q} + 744 + 196884 q + 21493760q^2 + 864299970q^3
         + 20245856256 q^4 +  \cdots
 \eeq
 while the Klein invariant has nice values at the parabolic and elliptic points of the
 fundamental domain of the modular group.
   Different definitions and normalizations
 of $\Delta$ exist in the literature, but all definitions of the $j$-functions
 are such that they lead to the same $q$-expansion.   The valence
 formula (\ref{valence-formula}) implies that the $j$-function is holomorphic on $\cH$,
  with a simple pole at infinity and a triple zero at $\xi_3=e^{2\pi i/3}$.
  This follows from the fact that the Ramanujan form $\Delta$ is holomorphic on $\cH$ with
   a simple zero at infinity, hence is non-vanishing
  on the upper halfplane. The Eisenstein series $E_4$ is homolorphic with
  a simple zero at $\xi_3=e^{2\pi i/3}$, hence is non-vanishing on $\cH$ and at infinity.

  The $j$-invariant is an ubiquitous function which plays an important
 role in completely different contexts.
 After having been discussed earlier by Kronecker in 1857 \cite{k1857}
  in the context of complex multiplication, and Hermite in 1858 \cite{h1858} in the
  context of solving the quintic equation, it was interpreted
   by Dedekind in 1877 \cite{d1877} as the function that
 maps the fundamental domain one-to-one onto $\mathC$.
  The $j$-function arises also  in the geometry of elliptic curves and in the representation
 theory of the Fischer-Griess monster group. The connection to the latter arises because
   its Fourier coefficients encode the dimensions of the representations of this finite group,
   thereby linking $j$ to the largest of the finite simple groups via the theory of
   vertex algebras \cite{b92}. Part of the importance of the $j$-function derives from the fact that every modular
function on the upper halfplane
 can be expressed as a rational function of $j$ \cite{kk98}.
   A brief history of this tantalizing function can be found in \cite{s01}.

  The Ramanujan form does not appear to be less versatile, with applications that
 range from a geometric interpretation as a motivic modular form, to  partition functions
  of the bosonic  string, as well as the entropy of certain types of black holes.

Inflationary potentials can be constructed by considering
dimensionless functions $F(j,\oj)$. An immediate class of examples
that can be considered is given by
  \beq
    V_p(\phi^1,\phi^2) ~=~ \L^4 |j(\tau)|^{2p}.
  \eeq
More general classes of models that could be
considered are based on potentials of the form
 \beq
   V_{p,q}(\phi^1,\phi^2) ~=~ \L^4 |j_q(\tau)|^{2p},
\eeq
 where $j_q(\tau)$ for $q$ prime are modular functions at level
 $q$ derived from $j$.

In the remainder of this paper the focus will be on the model with potential
 \beq
  V(\phi^1,\phi^2) ~=~ \L^4 |j(\phi)|^2.
  \eeq
 Figure 1 shows a graph of the absolute value of the $j$-function
  close to the boundary $(\rmIm ~\tau)=0$ of a domain $X$ that is
  centered around $(\rmRe~\tau)=0$.

\begin{center}
 \includegraphics[width=80mm,height=70mm]{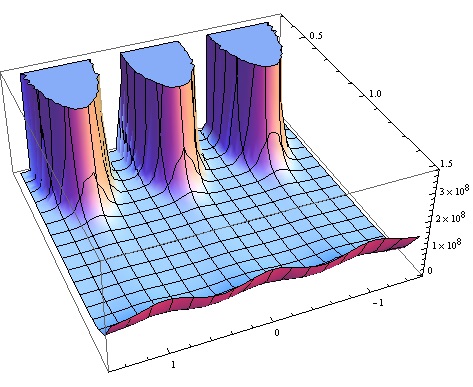}
\end{center}

\begin{quote}
 {\bf Figure 1.}~{\it The $j$-function close to the
 boundary $(\rmIm~\tau) =0$ in a region} \\
 {\it  of the upper halfplane $\cH$  centered around  $(\rmRe~\tau)=0$.}
 \end{quote}

 \vskip .3truein
\parindent=15pt

\section{Observables of $j$-inflation}

The satellite probes WMAP \cite{h12etal} and {\sc Planck}
\cite{planck15-13, planck15-20} have constrained the scalar power
spectrum, in particular its amplitude and spectral index, and put bounds on the tensor
amplitude. These results thus provide a set of observables $(A_{\cR\cR}, ~n_{\cR\cR}, ~r)$
for early universe models.

\subsection{Modular form of observables}

 In order to express the observables in terms of the basic functions of
  $j$-inflation, it is useful to write the slow-roll parameters $\e_I$ in terms
  of the Eisenstein series. The construction of the derivatives outlined above
  leads in the special case of the Eisenstein series
  to
     \bea
    E'_4 &=& \frac{2\pi i}{3} \left(E_4E_2-E_6\right) \nn \\
    E'_6 &=& \pi i \left(E_6E_2 - E_4^2\right),
  \eea
  relations that were originally obtained by Ramanujan \cite{r1916}.
  The appearance of the quasi-modular Eisenstein series $E_2$
  changes the transformation behavior
  of derivatives of modular forms.

  The observables of $j$-inflation can then be obtained by using in addition the 
  derivatives of the $j$-functions, which can be obtained as
   \bea
    \frac{j'}{j} &=& -2\pi i \frac{E_6}{E_4} \nn \\
    \frac{j''}{j} &=& -\frac{2\pi^2}{3} \left(4\frac{E_6^2}{E_4^2}
               + 3E_4- \frac{E_6E_2}{E_4}\right) .
  \eea
  This leads to the Eisenstein form of the slow-roll parameters
 \beq
  \e_I ~=~ -2\pi i^I \frac{M_\rmPl}{\mu}
       \left(\frac{E_6}{E_4} + (-1)^I \frac{\oE_6}{\oE_4}\right),
  \lleq{slow-roll-params}
  and the parameter that determines the slow-roll acceleration of
  the scale parameter $a(t)$ is given  by
  \beq
  \e_V ~=~ \frac{1}{2}G^{IJ} \e_I\e_J
    ~=~ 8\pi^2 (\rmIm ~\tau)^2 \frac{M_\rmPl^2}{\mu^2}
       \left|\frac{E_6}{E_4}\right|^2.
  \eeq
 This  implies that $\ddot{a}>0$ for values of $\tau$ that are close to the zero
  of the weight  six Eisenstein series, which is given by
  $\tau=i$. At the point  $\tau=i$ the Eisenstein series $E_4$ does not have a 
  zero or a  pole.

The  spectral index expressed in terms of the Eisenstein series then takes the form
 \beq
 n_{\cR\cR} ~=~ 1 - \frac{8\pi^2}{3} (\rmIm~\tau)^2 \frac{M_\rmPl^2}{\mu^2}
       \left[  8 \left|\frac{E_6}{E_4}\right|^2
             - 3 \rmRe \left( \frac{E_6\bE_4^2}{E_4\bE_6}\right)
                  +  \rmRe\left(\frac{E_6}{E_4}\bE_2\right)
      \right]
    ~+~ 8\pi \frac{M_\rmPl^2}{\mu^2} (\rmIm~\tau) \rmRe \left(\frac{E_6}{E_4}\right),
 \nn \\
\lleq{scalar-spectral-index}
\noindent
in dependence of $\tau =\phi/\mu$.
The last term on the rhs is induced by the curved target space metric.
It combines with the $E_2$-term in the square bracket to an expression that contains
the almost holomorphic modular form $\Ewhat_2$ as a factor
\beq
 n_{\cR\cR} ~=~ 1 - \frac{8\pi^2}{3} (\rmIm~\tau)^2 \frac{M_\rmPl^2}{\mu^2}
       \left[  8 \left|\frac{E_6}{E_4}\right|^2
             - 3 \rmRe \left( \frac{E_4^2}{E_6}\frac{\bE_6}{\bE_4}\right)
                  +  \rmRe\left(\Ewhat_2\frac{\bE_6}{\bE_4}\right)
      \right]
 \nn \\
\lleq{scalar-spectral-index-ahmi}
This can also be derived as a specialization of the general modular inflation result for
$n_{\cR\cR}$ derived earlier in the paper.

The amplitude $A_\cR$ of the scalar power spectrum can be expressed in
terms of the $j$-function and the Eisenstein series $E_w$ as
 \beq
 A_{\cR\cR} ~=~ \frac{1}{192\pi^4}
               \left(\frac{\L^2\mu}{M_\rmPl^3}\right)^2
     \frac{1}{(\rmIm~\tau)^2} \left|\frac{E_4}{E_6}\right|^2
     |j(\tau)|^2,
\lleq{scalar-amplitude-jinfl}
 where the rhs is to be evaluated with the inflaton values
$\tau^I=\phi^I/\mu$ such that $n_{\cR\cR}$ and the tensor-to-scalar ratio $r$, considered
below, are within
the experimental range. Once $ n_{\cR\cR}$ (and $r$) have
been used to determine $\tau_p^I$ at  the pivot scale one can use
the experimental result for $A_\cR$  to determine
the energy scale $\L$ of $j$-inflation.
The isocurvature power at horizon crossing is the same as that of the
 adiabatic perturbation $\cP_{\cS\cS} = \cP_{\cR\cR}$ and the
 cross correlation vanishes $\cP_{\cR\cS}=0$.

The tensor-to-scalar ratio $r$ of multifield inflation with
curved targets (\ref{tensor-ratio-mfi}) takes for modular inflation the
 form
 $
  r ~=~ 8(\rmIm~\tau)^2 \delta^{IJ}\e_I\e_J,
$
 and with the $j$-inflation expression for the parameters $\e_I$ one obtains  the
 Eisenstein form for $r$ as
  \beq
    r ~=~ 128 \pi^2 ~(\rmIm~\tau)^2 ~\frac{M_\rmPl^2}{\mu^2}
       ~\left|\frac{E_6}{E_4}\right|^2.
 \eeq
 In the lowest order in the slow-roll approximation this determines the tensor
 spectral index via (\ref{r-nt-relation}).

The evaluation of the observables is here at $t_*$, the time at which the pivot scale
crosses the horizon during inflation. The first constraint on the
parameters of the model can be obtained from the requirement that the
number of e-folds
 \beq
  N_* ~=~ \int_{t_*}^{t_e} dt~H(t)
 \eeq
 should fall into the standard range $N_* \in [60,70]$. This
 interval is not sharp, and values within a wider range have been
 considered in the literature. The input for the computation of
 the number of e-folds is the inflationary dynamics which in the slow-roll
 approximation takes for $j$-inflation the form
  \beq
   \phidot^I ~=~  \frac{2\pi i^I}{\sqrt{3}} \frac{1}{G}
   \left(\frac{M_\rmPl}{\mu}\right) \L^2
    \left(\frac{E_6}{E_4} + (-1)^I \frac{\bE_6}{\bE_4}\right) |j|,
 \lleq{sr-jinfl-dynamics}
 where $G=(\mu/\phi^2)^2$ is the conformal  factor of the modular
 inflation  target space metric. Integrating the $j$-function then leads
 to the number of e-folds
  \beq
   N_* ~=~ \frac{1}{\sqrt{3}} \frac{\L^2}{M_\rmPl} \
           \int_{t_*}^{t_e} dt ~|j(\tau)|.
  \eeq

\vskip .2truein

\subsection{$j$-inflation observables  and the {\sc Planck} probe}

The parameter space of $j$-inflation is stratified by the energy scale $\mu$
that enters all the observables. Given a specific choice of $\mu$
and the pivot value $\phi_*^I$ of the inflaton, the
spectral indices $n_{\cR\cR}$ and $n_T$, as well as the tensor
ratio $r$ can be computed from the Eisenstein formulae above.
Furthermore, the scale $\L$ can be
obtained via eq. (\ref{scalar-amplitude-jinfl}) from the
{\sc Planck} scalar amplitude \cite{planck15-13,  planck15-20},
which in turn allows to determine the number of e-folds $N_*$.

A detailed scan of the parameter space given by   $(\mu, \tau_*)$
 can be performed, leading to different neighborhoods
 $U_\mu(\tau_*)$ in the upper halfplane that can be tested
 against the satellite probe constraints. The form of the slow-roll
 parameters $\e_I$ in eq. (\ref{slow-roll-params}) and the
 scalar spectral index in eq. (\ref{scalar-spectral-index})
 show that the horizon crossing value of the inflaton should be chosen
 in a neighborhood of the zero of the Eisenstein series $E_6$,
 which can be obtained
  from the valence formula (\ref{valence-formula}) as $\tau = i=\sqrt{-1}$.
  The Eisenstein series $E_2$ and $E_4$
 are finite at this point \cite{kk98}
  \beq
   E_2(i) ~=~ \frac{3}{\pi},~~~~~
     E_4(i) ~=~ \frac{3\G(1/4)^8}{(2\pi)^6},
  \eeq
 hence $\e_I$, $n_{\cR\cR}$ and $r$
 are regular functions in this neighborhood. A zoom of the potential
 closer to this point with a particular inflaton trajectory
  is shown in Figure 2. It illustrates in more detail the
 ridge along the $(\rmRe~\tau)=0$ line that is suppressed in the large
 scale view of the potential of Figure 1.
 \parskip=0pt
 \begin{center}
\includegraphics[width=90mm,height=55mm]{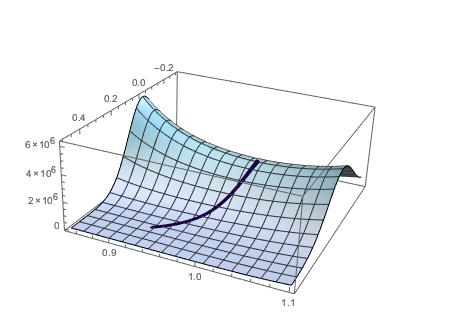}
\end{center}
\parskip=0pt
\begin{quote}
{\bf Figure 2.}~{\it A close-up of the $j$-inflation potential in the neighborhood
   of $\tau=i$.}   \\
  {\it The curved path indicates an $N_*=60$ trajectory.}
 \end{quote}

 \parskip=0.12truein
\vskip .2truein

\parindent=15pt
 After fixing the scale $\mu$ at a super-Planckian value,
 the inflaton values at horizon crossing can be chosen such
  that after integrating the $j$-inflation dynamics (\ref{sr-jinfl-dynamics})
  the orbit $\phi^I(t)$ leads to a number $N_*$ of e-folds between horizon crossing
  and the end of inflation that falls within  the standard range
  $
   N_* ~=~ [50,70].
  $
The field $(\phi^I)$ therefore traverses a super-Planckian distance in field space during
inflation. Figure 3 gives an illustration of the behavior for a few trajectories in the target space
$X$ associated to different scales $\mu$ that lead to the central value  $N_*=60$ and
are consistent with CMB phenomenology.
\begin{center}
\includegraphics[width=70mm,height=50mm]{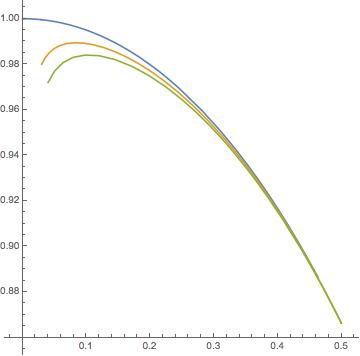}
\end{center}

\begin{center}
\begin{tabular}{l}
{\bf Figure 3.}~{\it Trajectories of the inflaton for
    different initial conditions with }  \\
  {\it   e-fold numbers $N_*$ in the interval $[50, 70]$.} \tabroom \\
\end{tabular}
\end{center}

These models in particular all lead to spectral indices
  $n_{\cR\cR} = 0.96*$,
 compatible with the {\sc Planck} result \cite{planck15-13}, and
 the tensor-to-scalar ratio takes values in the range
  $r \in  [10^{-8}, 0.08]$,
  compatible with  the result of the
  {\sc Planck} Collaboration \cite{planck13-16}, which reports for the tensor ratio $r$
  at the pivot scale  $k_p=0.002\rmMpc$ the bound $r_{0.002} \leq 0.11$,
  while the BICEP2/Keck/{\sc Planck}
  Collaboration reports $r_{0.05} \leq 0.12$ \cite{bkp15}.
 The energy scale $\L$ determining the height of potential
 includes the range
  $\L ~\in~ [10^{-6}, ~10^{-4}]M_\rmPl$
  for the realizations discussed here. More generally, the regions $U_\mu(\tau=i)$ around the
  slow-roll point $\tau=i$, identified above via the Eisenstein series, contain for varying $\mu$
   many orbits that are consistent with {\sc Planck} probe results. Moving too far from the slow-roll
   point violates the slow-roll condition, hence  leads to tensor ratios $r$ that are too large,
   as expected.
  On the inflationary time scales during which large scale perturbations cross the horizon
  the effect on the observables of  the transfer function $T_{\cO\cO'}$,
  which can be expressed in terms of the Eisenstein series $E_w$ using the results above
  and earlier in the paper, is small.

\vskip .2truein

\section{Conclusion}

Modular inflation is a class of two-field models obtained as a
specialization of multifield automorphic field theories by restricting the
automorphic group $G(\mathZ)$ to be given by the modular
group $\rmSL(2,\mathZ)$, or subgroups thereof.
The framework considered here differs from inflation
 theories based on moduli,  sometimes also called modular inflation. Moduli
inflation posits that some of the moduli that arise in string theory, in
particular in Calabi-Yau compactifications, are involved in the inflationary
process.  Cases of string theory induced potentials  exhibiting
automorphic symmetries  provide special examples that
fit into the framework of automorphic and modular inflation considered
here and in \cite{rs14, rs15}.

 Modular inflation models present the simplest class of theories
  that allow to embed the shift symmetry into a group, in the process
  leading to a stratified theory space, in which the individual leaves that
  provide the building blocks of the resulting foliation are
characterized by the weights and levels of the defining modular
forms. The field theory space of automorphic inflation in general, and
modular inflation in particular, has a nontrivial geometry that is encoded
in the Riemannian metric $G_{IJ}$ derived in a canonical way
from the underlying group structure. In this paper a detailed description
has been given of the general multifield curved target space dynamics
and its specializations to two-field inflation and
modular inflation, including a formulation of some
of the variables that enter the phenomenological analysis for
general modular potentials. An important problem arises from the fact that
derivatives of modular forms are not modular, raising the issue of the modular nature
of physical observables in modular invariant inflation. It was shown that
the nonmodular contributions of the derivatives of the inflaton potential combine
with the nonmodular terms induced by the curved target space into almost
holomorphic modular forms that in turn lead to CMB observables that are almost
holomorphic modular invariant.

An example of a modular function is given by the $j$-function,
defined as a quotient of modular forms of weight twelve forms
relative to the full modular group, i.e. level one. The
inflationary two-field model that results from the simplest
potential based on this function leads to a slow-roll
phenomenology that is consistent with the observational results
from the {\sc Planck} satellite probe. The realization of
$j$-inflation is constrained in particular by the bounds that have
been established in the past decade first by WMAP and more
recently by {\sc Planck} and the joint {\sc Planck}/BICEP collaboration,
in particular the bounds on the gravitational contribution to the power
spectrum.

\vskip .3truein

\parindent=0pt
{\bf \large { Acknowledgement.}} \hfill \break
 It is a pleasure to thank Monika Lynker for discussions. This
 work was supported by a Faculty Research Grant at Indiana
 University South Bend and benefited in the early stages from the
 hospitality of the Simons Institute for Geometry and Physics.

\end{document}